\documentclass[conference]{IEEEtran}
\IEEEoverridecommandlockouts
\usepackage{multirow}
\usepackage{multicol}
\usepackage[utf8]{inputenc} 
\usepackage{kotex} 

\usepackage{hyperref}
\usepackage{cite}
\usepackage{amsmath,amssymb,amsfonts}
\usepackage{algorithm}
\usepackage{algorithmicx}
\usepackage{algpseudocode}
\usepackage{graphicx}
\usepackage{textcomp}
\usepackage{xcolor}
\usepackage{xspace}
\usepackage{soul}
\usepackage{kotex}
\usepackage{booktabs} 
\usepackage{hyperref}
\definecolor{linkcolour}{rgb}{0,0.2,0.6}
\definecolor{xgreen}{rgb}{0.2,0.6,0.0}
\definecolor{xred}{rgb}{0.7,0.1,0.0}
\hypersetup{colorlinks,breaklinks,citecolor=red!50, urlcolor=linkcolour, linkcolor=xgreen}

\usepackage{amsmath}
\usepackage{color}
\usepackage{array}
\usepackage{stfloats}
\usepackage{amsthm} 
\usepackage{amssymb}
\usepackage{float}
\usepackage{nccmath}
\usepackage{graphicx}
\usepackage{setspace}
\usepackage[normalem]{ulem}
\usepackage{booktabs}
\usepackage{subcaption}
\usepackage{mwe}
\usepackage{cite}
\usepackage{graphicx}
\usepackage[export]{adjustbox}
\usepackage{amsmath,amssymb,amsfonts}
\usepackage{graphicx}
\usepackage{textcomp}
\usepackage{xcolor}
\usepackage[utf8]{inputenc} 
\usepackage{kotex} 

\usepackage{algorithm} 
\usepackage{algpseudocode}
\usepackage{grffile}

\usepackage{kotex}
\newcommand{\RN}[1]{%
  \textup{\uppercase\expandafter{\romannumeral#1}}%
}
\newcommand{\BfPara}[1]{\vspace{1mm}{\noindent\bf#1.}\xspace}

\def\BibTeX{{\rm B\kern-.05em{\sc i\kern-.025em b}\kern-.08em
    T\kern-.1667em\lower.7ex\hbox{E}\kern-.125emX}}
\begin{document}

\title{Quality-Aware Real-Time Augmented Reality Visualization under Delay Constraints}

\author{
\IEEEauthorblockN{Rhoan Lee}
\IEEEauthorblockA{\textit{Ewha Womans University} \\
Seoul, Korea \\
leerhoan@ewhain.net}
\and
\IEEEauthorblockN{Soohyun Park}
\IEEEauthorblockA{\textit{Korea University}\\
Seoul, Korea \\
soohyun828@korea.ac.kr}
\and
\IEEEauthorblockN{Soyi Jung}
\IEEEauthorblockA{\textit{Hallym University} \\
Chuncheon, Korea \\
sjung@hallym.ac.kr}
\and
\IEEEauthorblockN{Joongheon Kim}
\IEEEauthorblockA{\textit{Korea University}\\
Seoul, Korea \\
joongheon@korea.ac.kr}
}

\maketitle

\begin{abstract}
Augmented reality (AR) is one of emerging applications in modern multimedia systems research. Due to intensive time-consuming computations for AR visualization in mobile devices, quality-aware real-time computing under delay constraints is essentially required. Inspired by Lyapunov optimization framework, this paper proposes a time-average quality maximization method for the AR visualization under delay considerations. 
\end{abstract}

\section{Introduction}
In modern multimedia research trends, there have been a lot of attentions to augmented reality (AR) in terms of visualization, system designs, and network-based applications~\cite{icdcs21}.
In order to realize real-time AR visualization for various emerging applications in mobile and distributed devices, delay-aware methods are fundamentally required along with quality-aware visual computing~\cite{mm17,tmm2020long}.

In order to satisfy the delay requirements in real-time AR visualization, this paper proposes a novel quality-aware real-time AR visualization under the theory of Lyapunov optimization~\cite{ton201608kim}.
Based on the Lyapunov optimization, time-average utility maximization can be modeled under the consideration of queueing delays/stability. 
Therefore, this theory is widely applicable for the problems which deal with the tradeoff between utility maximization and stability, e.g., energy-delay tradeoff in networks~\cite{Ly1}, quality-delay tradeoff in streaming~\cite{tmc201907koo}, and accuracy-delay tradeoff in learning systems~\cite{Ly2}. 

\begin{figure}[t!]\centering
    \begin{multicols}{3}
        \includegraphics[width=0.97\columnwidth]{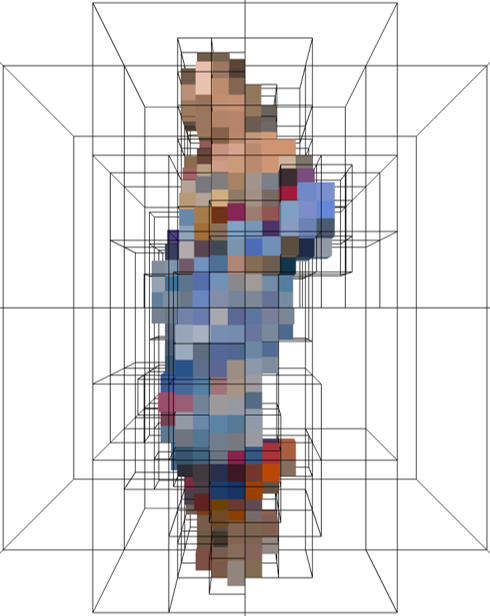}\captionsetup{justification=centering}
        \subcaption{Octree Depth: $5$ ($d=5$)}\label{normal_q}
        \includegraphics[width=0.97\columnwidth]{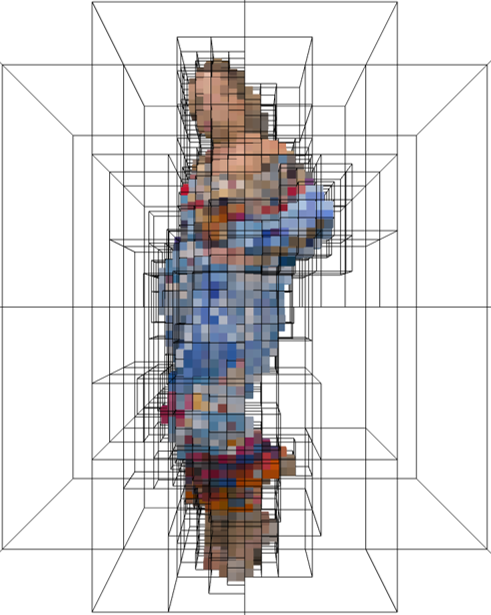}\captionsetup{justification=centering}
        \subcaption{Octree Depth: $6$ ($d=6$)}\label{heavy_q}
        \includegraphics[width=0.97\columnwidth]{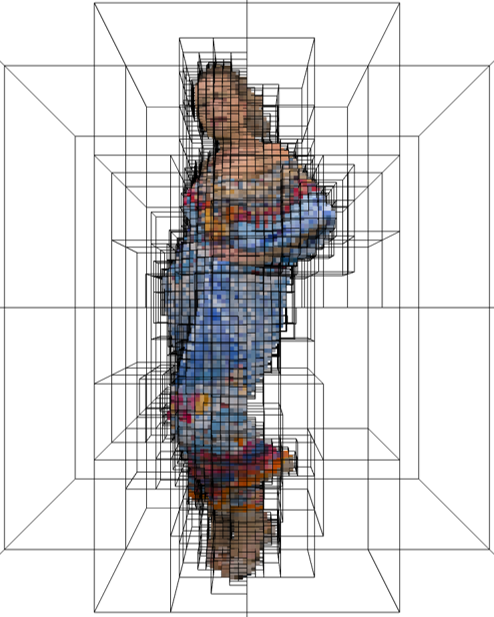}\captionsetup{justification=centering}
        \subcaption{Octree Depth: $7$ ($d=7$)}\label{normal_u}
    \end{multicols}
    \caption{AR visualization resolution depending on Octree depth.}%
    \label{fig:octree}%
    \vspace{-3mm}
\end{figure}

In real-time AR visualization, the quality of visualization can be controlled and quantifiable by Octree depth, as shown in Fig.~\ref{fig:octree}.  
Therefore, the proposed algorithm in this paper aims at the maximization of time-average AR visualization quality by controlling the Octree depth under stability. 
Furthermore, it has been shown that the proposed algorithm is also low-complexity and fully distributed. 
Through data-intensive performance evaluation, it has been confirmed that the proposed algorithm achieves desired performance improvements.

\vspace{-5mm}
\section{Quality-Aware AR Visualization}\label{sec:3}
\vspace{-5mm}
\BfPara{Algorithm Details}
We aim to determine the optimal Octree depths over time in order to maximize the time-average real-time quality for the AR visualization subject to rendering delays.
This can be formulated as,
\begin{eqnarray}
\max: & & \lim_{t\rightarrow\infty}\frac{1}{t}\sum_{\tau=0}^{t-1} P_{a}(d(\tau)) \\
\text{subject to} & & \lim_{t\rightarrow\infty}\frac{1}{t}\sum_{\tau=0}^{t-1} Q(\tau)< \infty \textrm{ (delay constraint)}
\label{eq:subjectto}
\end{eqnarray}
where $P_{a}(\tau)$ measures the quality of AR visualization with the Octree depth at $d(\tau)$, and $Q(\tau)$ represents the delays (i.e., AR streams that are ready to be visualized), over unit time $\tau$.

By increasing the depth of the Octree, better AR visualization quality can be achieved, whereas it introduces more computation time (i.e., delay). On the contrary, decreasing the depth results in a reduction of the quality of AR visualization, while it is beneficial due to low computation time (i.e., lower delays). 
In this application, Lyapunov optimization theory can be directly used to exploit this tradeoff and the following closed-form equation can be derived as~\cite{tmc201907koo}, 
\begin{equation}
    \boxed{d^{*}(t)\leftarrow
    \arg\max_{d(t)\in\mathcal{R}}
    \left[
        V\cdot p_{a}(d(t)) - Q(t)a(d(t))
     \right]},
\label{eq:lyapunov-AR}
\end{equation}
where $d^{*}(t)$, $\mathcal{R}$, $a(d(t))$, and $V$ stand for the optimal Octree depth decision at $t$, the set of Octree depth candidates, and the arrivals by the determined Octree depth $d(t)$, the tradeoff coefficient, respectively. 
Here, if we prioritize queue stability with a smaller $V$, it is clear that the algorithm operates to minimizes visualization delays for stabilization. Therefore, we can conclude that by controlling $d(t)$ in each time, we can maximize the time-average AR visualization quality subject to the queue stability by applying the formulation in \eqref{eq:lyapunov-AR}. 


\begin{algorithm}[t]
\caption{Our Proposed Stabilized AR Visualization}
\label{alg:dpp2}
\small
\begin{algorithmic}[1]
\Statex $\hspace{-1.5em}\textbf{Initialize:}$
\State $t\leftarrow 0$; $Q(t)\leftarrow 0$;
\State Decision Action: $\forall d(t)\in\mathcal{R}$ 
\Statex $\hspace{-1.5em}\textbf{Stabilized Adaptive Ship Detection:}$
\While{$t\leq T$} 
    \State Observe $Q(t)$;
    \State $\mathcal{I}^{*} \leftarrow \infty$;
    \For{$d(t)\in \mathcal{R}$}
        \State
        $\mathcal{I}
            \leftarrow V\cdot p_{a}(d(t)) - Q(t)a(d(t))$;
        \If {$\mathcal{I} \leq \mathcal{I}^{*}$}
            \State $\mathcal{I}^{*}\leftarrow\mathcal{I}$; $d^{*}(t)\leftarrow d(t)$;
        \EndIf
    \EndFor
\EndWhile
\end{algorithmic}
\end{algorithm}

\BfPara{Pseudo-Code and Complexity}
The pseudo-code of the proposed algorithm is presented in Algorithm~\ref{alg:dpp2}. 
From (line 1) to (line 3), system parameters are defined and initialized. In (line 5), the current delay $Q(t)$ is observed. From (line 7) to (line 12), the main procedure of the closed-form computation for \eqref{eq:lyapunov-AR} is executed. 
Since our proposed algorithm computes a closed-form equation with the number of decision actions, i.e., $|\mathcal{R}|$, the real-time computation is with low-complexity which is $O(N)$ where $N$ is the number of Octree depth candidates. In addition, our solution can be computed in a distributed manner, because it works with closed-form equation computation with no side information.

\section{Performance Evaluation}

\BfPara{Experiment Settings}
For evaluation, the performance differences among various Octree depth candidates are compared with 8i Voxelized full bodies point cloud dataset~\cite{7914676}. We adopt Open3D~\cite{zhou2018open3d} to implement point cloud reading, data format conversion, and Octree depth control functionalities.

\begin{figure}
    \begin{multicols}{2}
        \includegraphics[width=\linewidth]{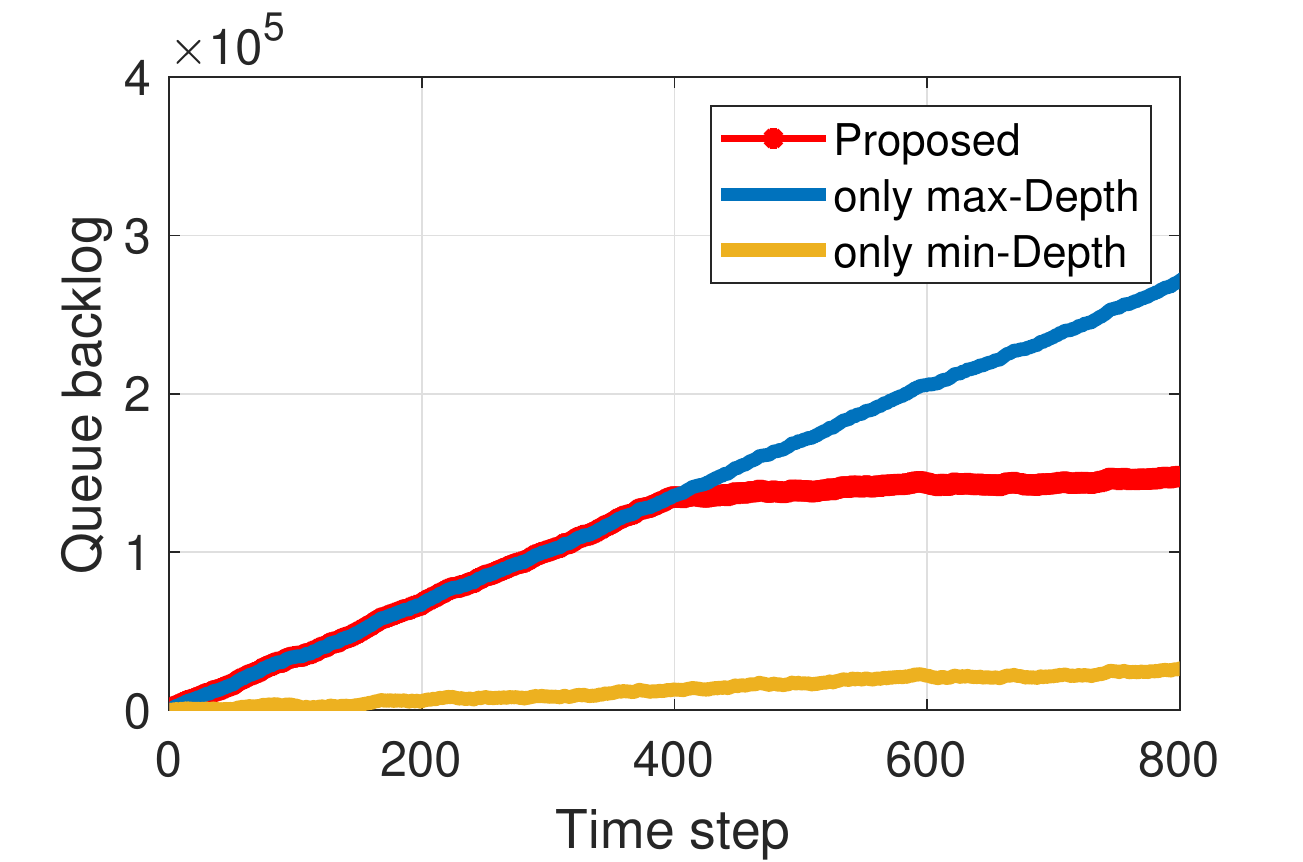}
        \subcaption{Queue/stability dynamics}\label{edge_pic}\par
        \includegraphics[width=\linewidth]{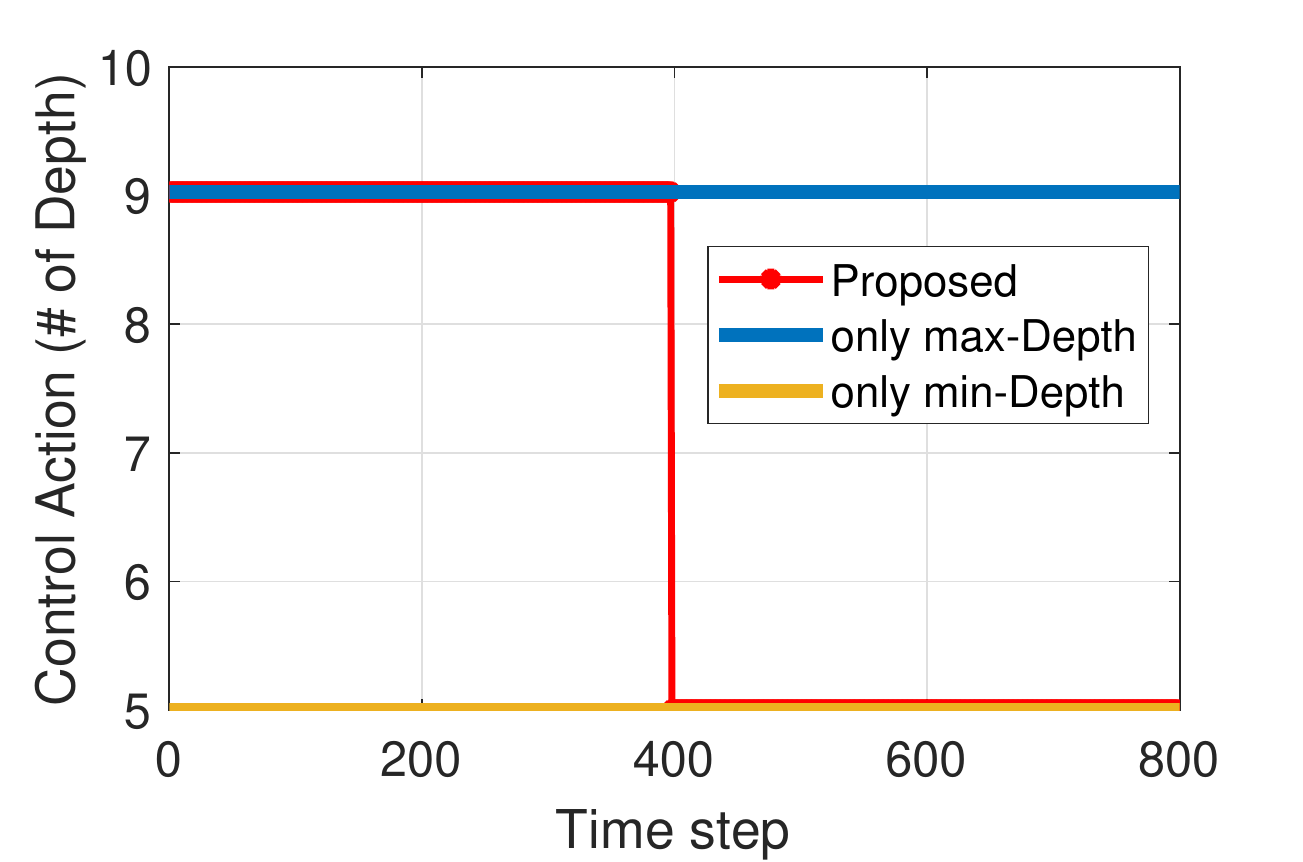}
        \subcaption{Control action updates}\label{heavy_q}\par
    \end{multicols}
    \caption{Performance evaluation results}%
    \label{fig:results}%
\end{figure}

\BfPara{Evaluation Results}
Fig.~\ref{fig:results}(a) shows the performance of the proposed method for time-average quality maximization by controlling Octree depths subject to delay conditions. Two controls, i.e., max-depth and min-depth, are used for optimal Octree quality-aware visualization performance. 
In general, larger the number of point clouds (PC) that is determined by the Octree depth obviously introduces higher AR visualization performance. Therefore, the bigger the number of PCs introduces better visualization quality; however longer delays with smaller the depth also introduces low-delay computation. 
As a result, the queue backlog size for only max-Depth diverges due to increased visualization computation delays and causes queue overflow after a certain time, while the stability of only min-Depth converges to $0$ due to fast visualization computation. Contrarily, the proposed scheme recognizes $400$ unit time as the optimized point and guarantees both delay stability and AR visualization quality.
Fig.~\ref{fig:results}(b) illustrates the performance of the proposed scheme under different control actions (i. e. the numbers of depth). The proposed method can ensure high-quality AR visualization until the recognized optimized point of $400$ unit time, where it drops in order to maintain given delay constraints. Consequently, we can conclude that the proposed algorithm provides maximized quality AR visualization while satisfying delay constraints, with Lyapunov optimization framework.

\section{Concluding Remarks}\label{sec:sec5}
In mobile devices, real-time AR is one of emerging applications in modern multimedia services. According to time-consuming computations for real-time AR visualization in mobile devices, quality-aware computing under delay constraints is essentially required. For this objective, this paper proposes a time-average quality maximization method for the AR visualization under delay considerations, inspired by Lyapunov optimization. Lastly, the performance was evaluated via data-intensive simulations.

\BfPara{Acknowledgment} This research was supported by NRF-Korea \& ITRC (2021R1A4A1030775 \& IITP-2022-2017-0-01637). S. Jung and J. Kim are the corresponding authors.

\bibliographystyle{IEEEtran}

\end{document}